\documentclass[fleqn,usenatbib]{mnras}

\usepackage{newtxtext,newtxmath}

\usepackage[T1]{fontenc}
\usepackage{ae,aecompl}

\usepackage{graphicx}	
\usepackage{amsmath}	
\usepackage{array}
\usepackage{newtxtext,newtxmath}

\title[ALMACAL-CO]{ALMACAL VIII: A pilot survey for untargeted extragalactic CO emission lines in deep ALMA calibration data}

\author[A. Hamanowicz et al.]{Aleksandra Hamanowicz,$^{1,2}$\thanks{E-mail: ahamanowicz@stsci.edu}
Martin A. Zwaan,$^{1}$, C\'eline P\'eroux,$^{1,3}$ 
\newauthor Claudia del P. Lagos,$^{4,5,6}$ Anne Klitsch,$^{7}$ 
 Rob J. Ivison,$^{1}$  Andrew D.~Biggs,$^{1}$
 \newauthor Roland Szakacs,$^{1}$ Alejandra Fresco$^{8}$ 
\\
$^{1}$ European Southern Observatory, Karl-Schwarzschild-Str. 2, 85748 Garching near Munich, Germany\\
$^{2}$ Space Telescope Science Institute, 3700 San Martin Drive, Baltimore, MD 21218, USA \\
$^{3}$ Aix Marseille Univ., CNRS, LAM, (Laboratoire d'Astrophysique de Marseille), UMR 7326, F-13388 Marseille, France \\
$^{4}$ International Centre for Radio Astronomy Research (ICRAR), M468, University of  Western Australia,\\ 35 Stirling Hwy, Crawley,WA 6009, Australia \\
$^{5}$ ARC Centre of Excellence for All Sky Astrophysics in 3 Dimensions (ASTRO 3D) \\
$^{6}$ Cosmic Dawn Center (DAWN), Niels Bohr Institute, University of Copenhagen, Lyngbyvej 2, 2100 Copenhagen, Denmark \\
$^{7}$ DARK, Niels Bohr Institute, University of Copenhagen, Jagtvej 128, 2200 Copenhagen, Denmark \\
$^{8}$ Max-Planck-Institut f\"ur extraterrestrische Physik (MPE), Giessenbachstrasse 1, D-85748 Garching bei M\"uchen, Germany \\
}

\date{Accepted XXX. Received YYY; in original form ZZZ}

\pubyear{xxx}

\begin{document}
\label{firstpage}
\pagerange{\pageref{firstpage}--\pageref{lastpage}}
\maketitle

\begin{abstract}
We present a pilot, untargeted extragalactic carbon monoxide (CO) emission-line survey using ALMACAL, a project utilizing ALMA calibration data for scientific purposes. In 33 deep ($T_{\rm exp}$ > 40~min) ALMACAL fields we report six CO emission-line detections above $S/N$ > 4, one third confirmed by MUSE observations. With this pilot survey, we probe a cosmologically significant volume of $\sim 10^5$~cMpc$^3$, widely distributed over many pointings in the southern sky, making the survey largely insusceptible to the effects of cosmic variance. We derive the redshift probability of the CO detections using probability functions from the \textsc{Shark} semi-analytical model of galaxy formation. By assuming typical CO excitations for the detections, we put constraints on the cosmic molecular gas mass density evolution over the redshift range $0 < z < 1.5$. The results of our pilot survey are consistent with the findings of other untargeted emission-line surveys and the theoretical model predictions and currently cannot rule out a non-evolving molecular gas mass density. Our study demonstrates the potential of using ALMA calibrator fields as a multi-sightline untargeted CO emission line survey. Applying this approach to the full ALMACAL database will provide an accurate, free of cosmic variance, measurement of the molecular luminosity function as a function of redshift. 

\end{abstract}



\section{Introduction}
  
The cosmic star-formation history \citep[SFH,][]{madau14} establishes the peak of the star formation in the Universe to be two billion years after the Big Bang ($z \sim$ 2), followed by order of magnitude decline to the present day. To understand what drives the cosmic SFH, we need to look at how the elements that lead to star formation evolve throughout cosmic time. 

Hydrogen is the most abundant element in the Universe and, in its different phases, is an ideal tracer of the baryon cycle - from its ionised state in the intergalactic medium to its neutral phase in the interstellar medium of galaxies \citep[see the review of][]{peroux20}. Moreover, several studies have established that hydrogen in its molecular form is a direct fuel for star formation, as it is the gas phase that is most tightly correlated to the star-formation rate in galaxies \citep[see review by][]{krumholz14}. Hence, to answer why the cosmic SFR evolves with the characteristics described above, it is essential to study how the molecular gas content of galaxies and the Universe evolves. 

Direct measurements of the molecular gas mass density $\Omega_{\rm H_2}$ through H$_{\rm 2}$ emission are, however, impossible for extragalactic sources as this molecule is chracterised by faint vibrational transitions \citep[e.g.][]{cui2005}. Instead, the molecular gas content of galaxies largely relies on observations of the second most abundant molecule, CO, whose bright transitions are observable out to the distant Universe \citep[$z > 2$,][]{carilli, hodge20}. CO has been detected in star-forming galaxies at different redshifts through targeted surveys \citep[e.g.][]{greve05, daddi10,genzel10, bothwell13, freundlich19, tacconi20}, providing us with the view of the molecular gas reservoirs in massive galaxies and their link to star formation. Such surveys target massive star-forming galaxies and have been used to investigate the scaling relations linking the galaxies' molecular gas content with other galaxy properties such as stellar mass or SFR. Extrapolating these scaling relations to lower mass regimes provides an approximate overview of the molecular gas content of galaxies at different redshifts.

Although powerful, scaling relations from targeted surveys likely introduce unknown systematic biases in the measurements of $\Omega_{\rm H_2}$. An unbiased way to study the molecular gas content of galaxies across cosmic time is through untargeted emission-line surveys: observation over a selected sky area without target pre-selection. So far, untargeted CO surveys have focused on cosmological fields with a significant multi-wavelength coverage. ASPECS \citep{walter16, aspecs} in HUDF provided robust constraints on the molecular gas mass function evolution up to redshift $z = 4$ and combined the CO detections with the HUDF optical counterparts. COLDz in the COSMOS and GOODS-North fields \citep{coldz} added the molecular gas mass function constraints at the redshift range $5 - 7$ and quantified CO luminosity functions at median redshfit of $<z> = 2.4$. However, these surveys have been limited to single fields, covering a small part of the sky, making them quite susceptible to the possible effects of cosmic variance. Additionally, \citet{phibbs2} searched for secondary sources in the multiple NOEMA PHIBSS2 survey fields, further supported the measurements reported by the ASPECS group. (originally survey targeting CO emission in massive star-forming galaxies). 

To quantify the molecular gas budget of the Universe and the impact of the molecular gas content of galaxies on the Universe's SFH, we need a statistically significant sample of well-characterised CO-selected galaxies. Such a sample can be provided by ALMACAL: an untargeted survey using ALMA calibration data obtained semi-randomly across the southern sky. Before embarking on a comprehensive untargeted molecular emission-line search on the complete ALMACAL data set, we decided first to run a pilot, proof-of-concept, ALMACAL-CO survey. Testing the survey concept on a selected dataset provides a training set on which we can explore the systematics and biases of the ALMACAL data. We are also introducing a novel statistical approach, addressing the emission-line classification challenges of a sky-wide survey without corresponding deep optical follow-up observations. We construct the redshift probability functions of each emission-line detection based on the CO flux predictions from the \textsc{Shark} Semi-Analytical Model of galaxy formation \citep{lagos18, lagos20}. Additionally, we complement these estimates using an empirical classification of the detections as the lowest possible CO transition observable at a given frequency. We emphasise that the results presented in this paper are preliminary and exploratory; with the future main survey spanning over all ALMA calibrator fields, we will significantly improve the statistics and be able to provide stringent constraints on the CO luminosity functions as well as the evolution of the molecular gas mass density with redshift. 

This paper is organized as follows: Section 2 describes the data and the reduction process; Section 3 presents the source selection. Molecular gas mass density calculations are presented in section 4. We discuss our findings, comparing them to previous studies in section 5, and we summarise the results in Section 6. We adopt the following cosmology: $H_{0}$ = 70 km s$^{-1}$ Mpc$^{-1}$, $\Omega_{\rm M}$ = 0.3, $\Omega_{\rm \Lambda}$  = 0.7.

\begin{figure}
	\includegraphics[width=\columnwidth]{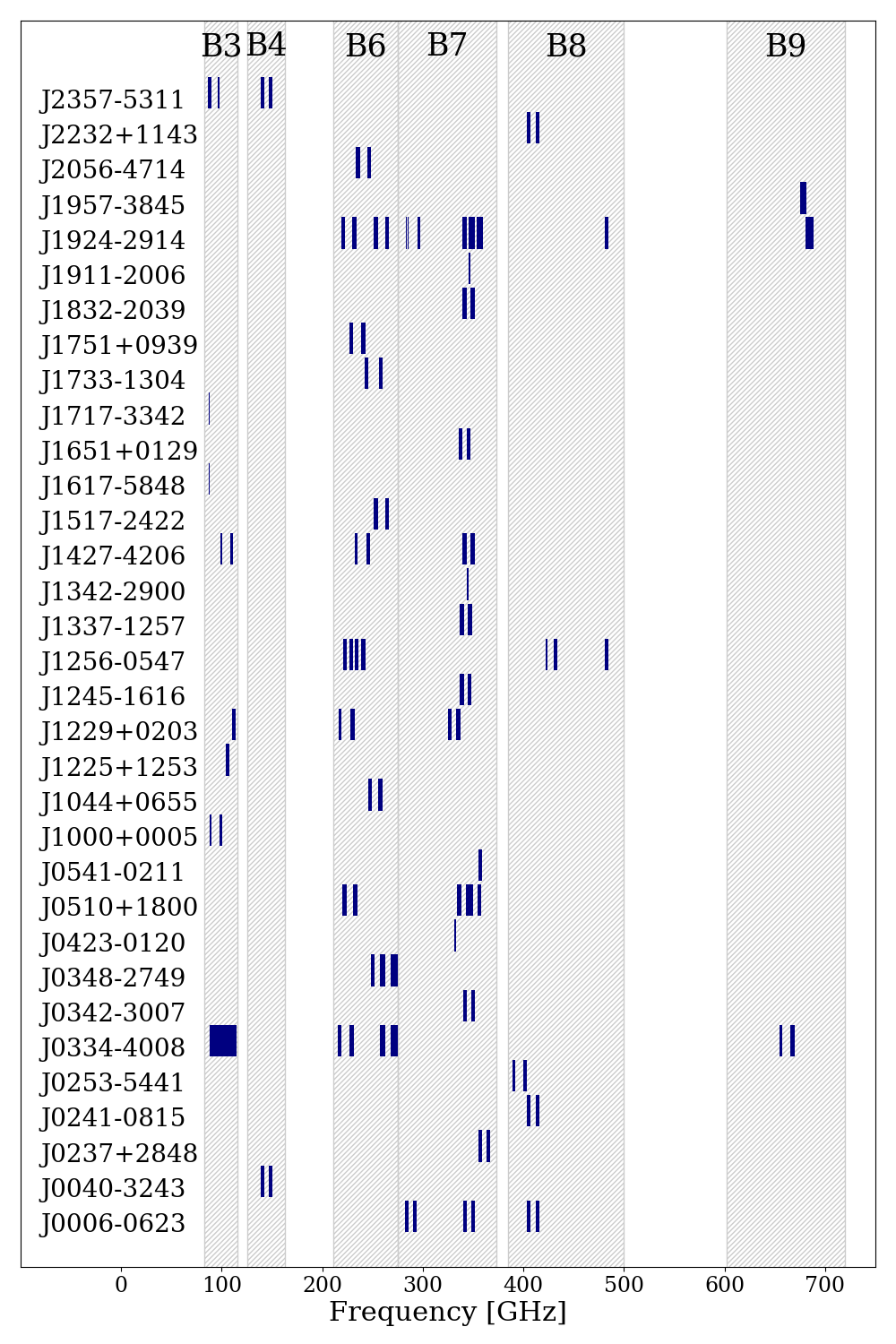}
    \caption{Frequency coverage of ALMACAL-CO pilot deep fields (T$_{\rm int} > $ 40 min). The fields are presented with increasing right ascension from bottom to top. Blue rectangles mark the frequency coverage of the data used in this study. Grey shaded areas represent the ALMA bands' frequency coverage labeled on the top. The frequency coverage of ALMACAL observations depends on the PI's requests for corresponding ALMA science projects. The coverage of different calibrators depends on their popularity, mainly the position in the sky near the most often observed targets.}
    \label{fig:frequency}
\end{figure}

\section{An untargeted emission-line survey}
\subsection{The ALMACAL survey}

The ALMACAL project\footnote{\url{almacal.wordpress.com}} is utilising the archival observations of ALMA calibrators for scientific purposes \citep{oteo16}. Each scientific observation with radio and millimetre interferometry requires several short integrations on a calibrator located at a small separation from the science field. The calibrator is observed with a short exposure time (several minutes) with a setup identical to the science observations requested by the project's PI (spectral resolution and coverage). In addition, less frequent but deeper integrations of bandpass calibrators are taken. The frequency coverage and depth of the observations depend on the popularity of the calibrator. Frequency coverage of the ALMACAL fields included in this study can be found in Fig. \ref{fig:frequency}.

ALMA calibrators are bright sub-millimetre point sources distributed over the sky accessible to the observatory. The presence of luminous sources in the centre of the field of view may raise concerns about possible object clustering around these sources. This issue was already investigated by \citet{oteo16} in their untargeted survey for submillimetre galaxies (SMGs) in ALMACAL fields. The vast majority of calibrators in our sample are classified as blazars \citep{bonato18}, which are bright sub-mm sources because of their orientation \citep[jet pointing towards the observer][]{urry95}. Nevertheless, every galaxy survey conducted on the ALMACAL data may be minimally biased towards over-densities, especially for sources/calibrators with unknown redshift.

According to ALMA policies, ALMA calibration scans become publicly available after the data sets that include these scans have passed the final quality assurance steps and have been delivered to the principal investigator. For every new observation, the calibrator scans are delivered together with the science data.

We extract data from the ALMA archive, remove the science observations and reduce the calibration data separately. The data are self-calibrated on the bright central source with a custom-made pipeline, finding complex gain solutions in the shortest time intervals allowed by the data. The bright point sources are modelled and subtracted from the visibility data, leaving calibrator-free data sets \citep[see][for detailed descriptions of the process]{oteo16}. Finally, all data are re-sampled to the exact spectral resolution. An extensive range characterises the raw data used by ALMACAL in spectral resolution. The original ALMA data are either in so-called FDM (frequency division mode), which means high spectral resolution, or TDM (time division mode), which is low resolution or continuum mode. To homogenize the data and reduce the total data volume, all FDM data are reduced to the lowest available channel separation of 15.6~MHz, similar to the resolution of TDM data. By default, the correlator software applies Hanning smoothing to all ALMA data.

This implies that data taken in TDM mode have an intrinsic resolution of 31.2~MHz. However, the spectrally averaged data to the TDM resolution in the ALMACAL pipeline have a resolution closer to 15.6~MHz. These are the majority of the data sets. Since different data sets can be combined to produce cubes, the typical final resolution of the data cubes is larger than 15.6~MHz, while the channel separation is always exactly 15.6~MHz (which corresponds to 46.8 km s$^{-1}$ in Band 3, 14.3 km s$^{-1}$ in Band 6 and 10 km s$^{-1}$ in Band 8). 

The ALMACAL database includes observations from all ALMA bands, most of which are in Band 3 ($84 - 116$~GHz) and Band 6 ($211 - 275$~GHz). Up until today, the ALMACAL database consists of over 2500 hours of observations (the equivalent of about one-half of a full ALMA yearly observing cycle) and around 1000 calibrator fields. We are accumulating new data constantly, but new fields are added only sporadically. 

\subsection{ALMACAL-CO sample selection}
\begin{figure}
	\includegraphics[width=\columnwidth]{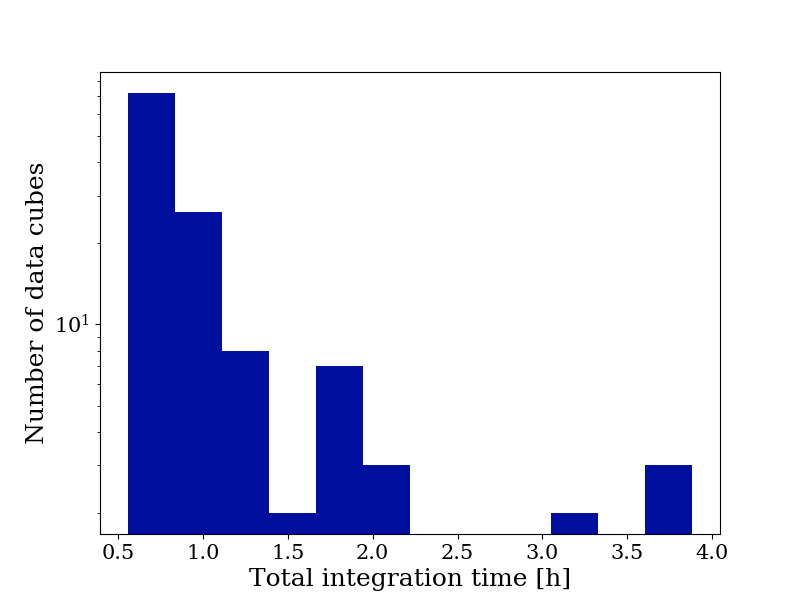}
    \caption{Total integration time distribution of the ALMA data cubes used in this study. The lowest integration time is limited to 40 minutes by construction. The longest integration time typically reaches 2 hours with a few exceptions of even longer observations. The total integrated time results from concatenating the short (several minutes) singular calibration pointing taken in the same spectral setup. }
    \label{fig:time}
\end{figure}

ALMACAL-CO is a project aiming at measuring the evolution of the cosmic molecular gas mass function through the untargeted detections of CO emission lines. This work presents a pilot project focusing on the fields with the longest integration times, which we refer to as \textit{deep}. The data selection was performed in September 2017. Below we describe the selection of the data for the pilot study.

\begin{figure}
	\includegraphics[width=\columnwidth]{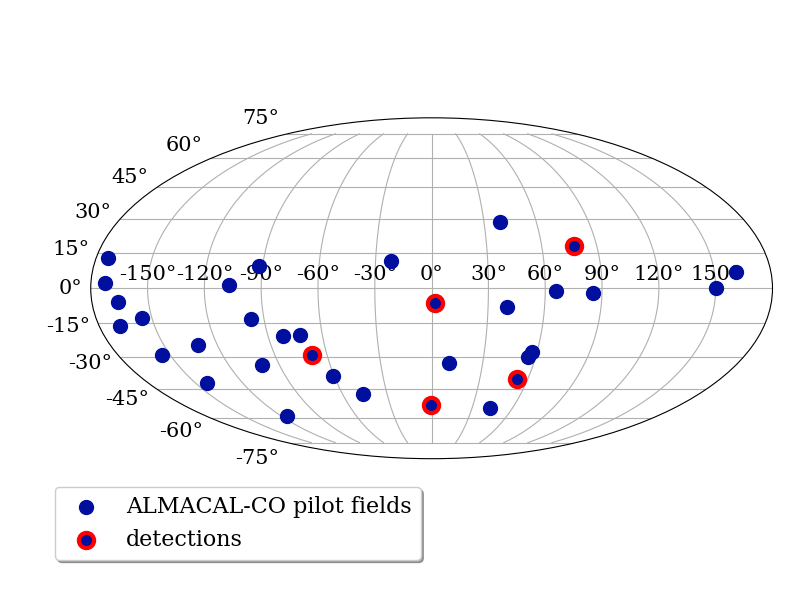}
    \caption{The ALMACAL-CO deep field distribution on the sky (blue circles). Red circles mark five fields, with six reported detections included in this study. The size of the points does not represent the physical scale. The random distribution of fields over the sky makes the survey less prone to the effects of cosmic variance. }
    \label{fig:sky}
\end{figure}

For each ALMACAL calibrator field, we chose observations undertaken in the same frequency setup accumulating a total observation time longer than 40 minutes. To ensure that the frequency coverage of such selected observations are identical, we match the calibrator observations by the proposal IDs and  check that the effective frequency coverage is identical. In this way we exclude spectral scans, or observations with overlapping but not identical frequency coverage. The cubes vary in depth (integrated time) from 40 minutes to 4 hours (Fig. \ref{fig:time}). The chosen observations were concatenated for each field into a single file and converted to a data cube with standard \textsc{CASA} routines. Before concatenation, we imaged each observation and visually checked the quality to remove any data suffering from artefacts related to imperfect calibration or dynamic range limitations. The data selection resulted in 147 cubes over 37 calibration fields, each cube representing a different frequency coverage.  
The bright calibrator at the centre of each ALMACAL field is removed before our analysis. However, if the original object has distinct spectral features (e.g. broad lines, prominent continuum slope) this automated procedure can leave residuals in the cubes. To account for that, we ran a continuum fitting routine (\textsc{uvcontsub}) on the concatenated files and removed large-scale noise structures. Nevertheless, 10 per cent of the sample was still affected by imperfect calibrator subtraction and these affected cubes were removed from the analysis. The final sample consists of 133 cubes. The distribution of the fields in the sky is presented in Fig. \ref{fig:sky}.

The spatial extent of each image cube is set by the primary beam size, which depends on the observing frequency. The full width at half maximum (FWHM) of the primary beam ranges from 56~arcsec in Band 3, 27~arcsec in Band 6, to 9~arcsec in Band 9. We used the \textsc{CASA Analysis Utilities} package's routines to estimate the best parameters for constructing the images: the size of the pixels and the number of pixels across the image. In constructing the cubes, we combine data with an extensive range of spatial resolutions. To produce a dataset with consistent properties, which allows optimal detection of unresolved line emission, we taper all visibility data to the exact spatial resolution of 0.5~arcsec, limiting the minimal pixel size to 0.17~arcsec. In combining the data, we exclude spectral windows narrower than 0.5~GHz. In this project, we also do not consider observations from the Atacama Compact Array array.

After creating the data cubes, we calculated each cube's total root-mean-square (r.m.s.) and created noise-per-channel diagrams. The resulting analysis showed that some ranges of channels extend over the mean noise levels and revealed highly variable noise patterns throughout the cube. Therefore, we removed data cubes from the sample for which the r.m.s. was higher than 5~mJy beam$^{-1}$. Additionally, we flagged the channels with noise levels averaged over the field-of-view exceeding 1$\sigma$ of the noise variation in the cube (less than 20 per cent of all data). 

Because of the nature of the ALMACAL data, the spectral coverage of each observation depends on the spectral setting defined by the PI of the corresponding science project (Fig. \ref{fig:frequency}). The non-uniform spectral coverage of our fields results in a variation of the probed volume (see Table \ref{tab:volumes}. The inhomogeneity of the frequency coverage is taken into account in our volume calculations (Section \ref{sec:volumes}). In some of the calibrator fields, the frequency coverage is scarce, making the untargeted detection of the CO relatively unlikely.


\section{Untargeted CO line search}

\subsection{Source detection}

The untargeted emission-line search over ALMA data cubes in the ALMACAL-CO pilot sample was performed using the SoFiA open-access source finder \citep{sofia}. SoFiA is a flexible source finder developed primarily for searching and characterizing detections from large HI 21-cm surveys. The source finder searches the 3D data cubes on multiple scales, both spectrally and spatially, allowing for different source-finding algorithms, including classical sigma clipping or more advanced wavelet reconstruction. In addition, it allows for customised treatment of the noise: depending on the cube parameters, noise can be calculated globally or in each channel separately. 

We searched for the emission lines in the cubes before applying the primary-beam correction (which is required to perform flux measurements). The source finding was performed using the sigma clipping algorithm and spectral smoothing over a set of kernels increasing spatially and spectrally. This method allowed a fast analysis of the cubes in only a few minutes each. We limited the search to a 3$\sigma$ threshold and required the minimal size of the detection to be two pixels (0.3~arcsec) and three channels (corresponding to 140 km s$^{-1}$ in Band 3, 43 km s$^{-1}$ in Band 6 and 30 km s$^{-1}$ in Band 8). 

We devised several criteria for initial detections to be considered as candidates. They should have a peak flux signal-to-noise > 4 and reliability $R > 0.3 $ (Sec. \ref{sec:reliability}). Additionally, as we searched the non-primary-beam corrected cube, we removed candidates detected in regions where the primary beam gain is less than 0.3. At this level, we retrieve less than 30 per cent of the original flux, making the candidates from this region unreliable. We also reject the emission lines of which the width is larger than a quarter of the spectral extent of the cube. Lastly, we inspect visually all the detections and their positions to make sure that the candidate is not confused with the injected mock sources (Sec. \ref{sec:completeness}), which escaped our matching pipeline, or is the result of incorrect pixel merging of the final detection by the source finder. The properties of the candidates are summarised in Table \ref{tab:candidates}, while spectra and moment maps are shown in Fig. \ref{fig:det}.

\begin{figure*}
\includegraphics[width=2\columnwidth]{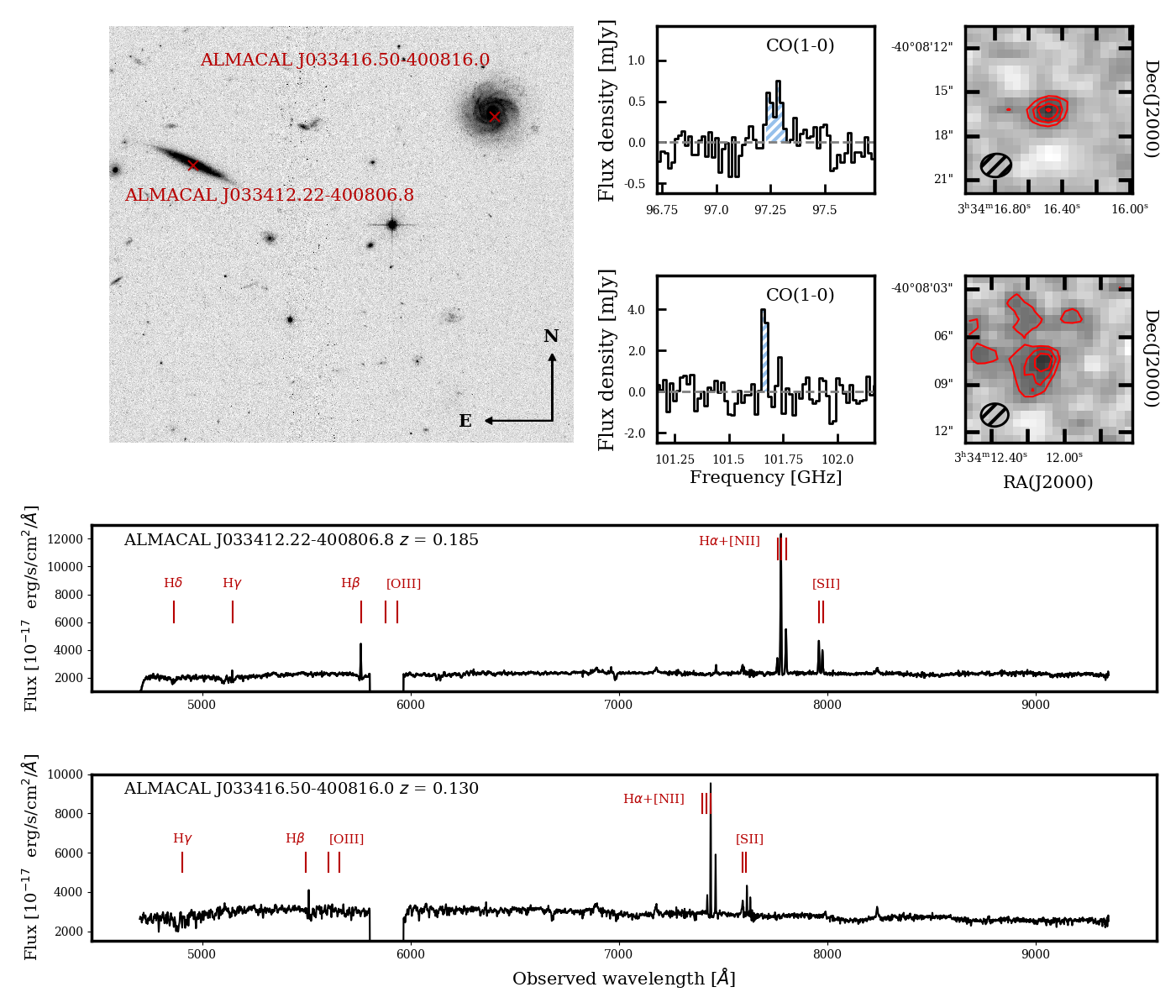}
\caption{Combination of optical (\textit{HST} imaging, MUSE spectroscopy) and ALMA observations of the ALMACAL-CO field J0334-4008. Two detected emission lines in that field coincide spatially with the two spiral galaxies, edge-on at $z = 0.185$ and face-on at $z=0.133$, allowing for the direct identification of the lines. The top left panel shows the \textit{HST} image of the field with the positions of ALMACAL-CO detections marked with red crosses. The two top right panels show the CO(1$-$0) emission-line detections for each galaxy and flux maps centred on the ALMA detections with $+$3,4,5,6 $\sigma$ contours. The bottom panels show MUSE spectra of the host galaxies, with the prominent emission lines marked in red. The gap in the spectra around 5800~\AA\  arises from the use of the Adaptive Optics system for these observations. }
\label{fig:j0334}
\end{figure*}

\begin{figure*}
     	\includegraphics[width=\textwidth]{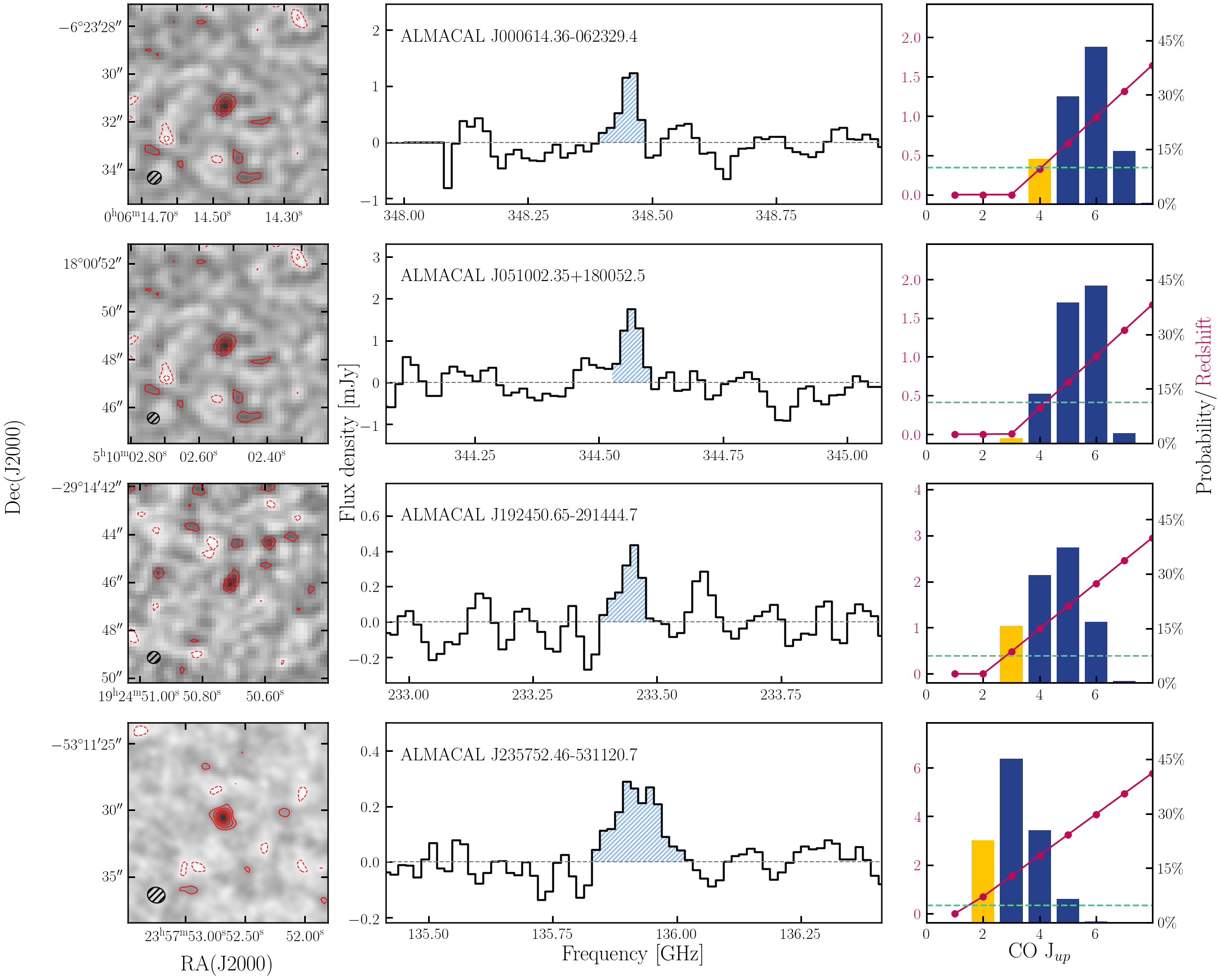}
     \caption{ALMACAL-CO emission lines candidates. Left: ALMA 0th-moment map centred on the detection candidates discovered in the ALMACAL-CO untargeted search. The contours correspond to $\pm$3,4,5,6$\sigma$, derived from the respective line map. The size of the synthesized beam is shown in the lower-left corner. The grey areas mark the parts of the primary beam that were not included in our image cut.
     Middle: spectrum of the line candidates at the brightest pixel. The blue dashed region marks the spectral extend of the detected line. Right: The distribution of CO transitions predicted by \textsc{Shark} corresponding to ALMACAL-CO line detections. The majority of sources in our sample are associated with $J_{\rm up}$ = 4,5,6. The parameters of all candidates as summarised in Table \ref{tab:candidates}. The lowest possible J$_{up}$ is marked in yellow. The read line indicates the redshift (left $y$-axis) of the corresponding $J$-transition in the histogram.The green horizontal dashed line marks the redshift of the calibrator in that field, indicating that the emitters do not preferentially cluster around the calibrator. For the candidates from field J0334-4008 (Fig. \ref{fig:j0334}), we choose the CO(1$-$0) solution with probability 100 per cent, as confirmed by the optical counterparts.}
     \label{fig:det}
 \end{figure*}

Deep ancillary data, which would ease the identification of the detections through optical counterparts, are mostly unavailable for our calibrator fields. However, for one of the fields in which we report two candidate detections, archival VLT/MUSE and HST data are available. Therefore, we identify the optical counterparts for these two CO detections in field J0334-4008 and confirm the redshifts (Fig. \ref{fig:j0334}). Although this does not guarantee the total reliability of our candidate sample, it demonstrates that our method is viable and that ALMACAL can be employed to make untargeted extragalactic CO emission-line detections.   

To test if our candidate detections are consistent with a random spatial distribution, we compared the distribution of the distances of the mock sources from the centre (which were injected at random positions, section \ref{sec:completeness}) to the distribution of the retrieved mock sources. A K-S test resulted in a $p$-value of 0.007, confirming that the spatial distribution of the retrieved sources is consistent with that of the randomly injected ones. 
Additionally, we checked the positions of the six candidate detections - although they are preferentially found beyond the half-power beam region, they do not appear to be detected at any preferential position. We conclude then that the detections are not biased towards a given position in the cube.

The ALMACAL-CO pilot survey is complementary to other untargeted CO emission line surveys conducted so far, in its strategy, spatial and frequency coverage. The survey includes coverage at higher frequencies, and is therefore sensitive to high-$J$ CO transitions around redshift $z = 0.5-1.5$ and [CII] at $z = 4-5$. 

\begin{figure}

	\includegraphics[width=\columnwidth]{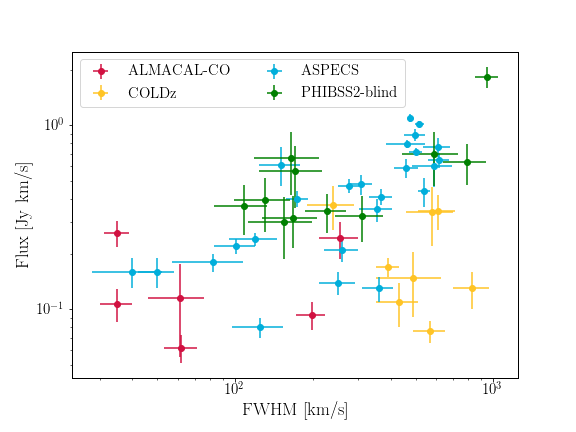}
    \caption{The integrated flux and width of the CO candidates from ALMACAL-CO (red) compared to detections from other untargeted surveys: ASPECS \citep[blue,][]{aspecs}, COLDz \citep[yellow, ][]{coldz}, PHIBSS2 \citep[green,][]{phibbs2}. ALMACAL-CO detections are fainter and narrower compared to the detections from other surveys. }
    \label{fig:flux}
\end{figure}


\subsection{Completeness}
\label{sec:completeness}

\begin{figure}

	\includegraphics[width=\columnwidth]{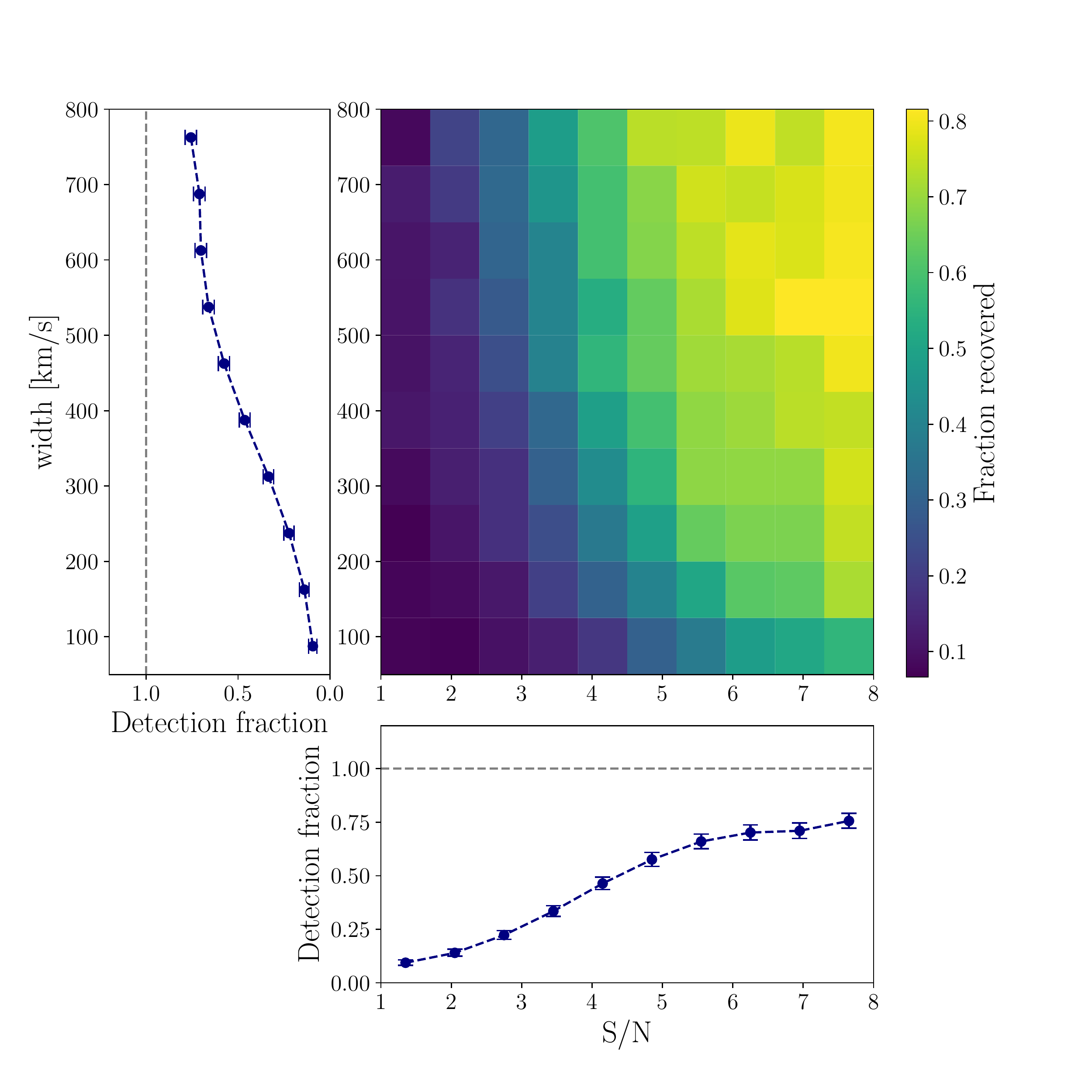}
    \caption{Completeness of the SoFiA search for line emitters in ALMACAL-CO pilot cubes. The detection fraction of line emitters as a function of their peak signal-to-noise ratio as well as the signal width in km s$^{\rm -1}$ is represented by colours on the 2D histogram. The additional panels represent the detection fraction as a function of one of the emission line parameters. The gradient in retrieving the unresolved mock sources points toward strong dependence of the success on the emission line velocity width as opposed to signal-to-noise. The highest completion rate has broad and strong signals (top right corner of the plot).}
    \label{fig:completeness}
\end{figure}

 \begin{table*}
	\centering
	\caption{The completeness and reliability coefficients (with errors) of ALMACAL-CO emission-line candidates. The columns: (1) Candidate ID including sky coordinates of the detection in J2000, (2) completeness (3) reliability coefficient. The two detections in the J0334-4008 field (last two rows) were confirmed through the MUSE observations, therefore we assign them with the reliability of 1.00.}
	\label{tab:coeff}
	\renewcommand{\arraystretch}{1.5}
	\begin{tabular}{lcc} 
		\hline \hline
		Source &  Completeness & Reliability   \\

		\hline 
ALMACAL J000614.36$-$062329.4 &	0.48 $^{+0.03}_{-0.02}$ & 0.51 $\pm$ 0.05 \\
ALMACAL J051002.35$+$180052.5 &	0.29 $^{+0.03}_{-0.02}$ & 0.32 $\pm$ 0.02 \\
ALMACAL J192450.65$-$291444.7 &	0.29 $^{+0.03}_{-0.02}$ & 0.38 $\pm$ 0.04 \\
ALMACAL J235752.46$-$531120.7 &	0.63 $^{+0.04}_{-0.03}$ & 0.74 $\pm$ 0.08 \\
\hline
ALMACAL J033416.50$-$400816.0 & 0.67 $^{+0.03}_{-0.03}$ & 1.00 $\pm$ 0.00 \\
ALMACAL J033412.22$-$400806.8 &	0.48 $^{+0.03}_{-0.02}$ & 1.00 $\pm$ 0.00 \\   
\hline
\hline
	\end{tabular}
\end{table*}

To measure the completeness of ALMACAL-CO, we inject mock sources into the survey's data cubes and feed them to the SoFiA source finder with the same setup as the actual search procedure. We inject 20 mock sources per cube, and repeat the random mock source injection 20 times in each cube of the survey to populate the mock sources parameter space. 

Mock sources are designed to mimic real emission lines and are described by several parameters: peak flux, central frequency, and width. For each source we assume a gaussian line profile with a random central frequency within the spectral range of a single cube. The peak flux was defined to have a range of 1 to 8 times the r.m.s. (in the steps of 0.1) of the noise in the given cube. The detection width was then assigned randomly from a uniform distribution between 10 and 500 km s$^{\rm-1}$ . Finally, we assign each detection a random spatial position in the cube, excluding a three-pixel-wide frame around the edges. After each injection, the positions become forbidden for other ingestions within a $\pm$ 3-pixel radius (the equivalent of the beam size) to prevent overlapping sources. All mock sources are spatially unresolved. 

We perform the source search with the SoFiA source finder to retrieve mock sources and detect real sources in the same search run. We compare the position of the detections with the injected mock sources catalogue to flag real candidates. 

The completeness as a function of emission-line strength and width is shown in Fig. \ref{fig:completeness} and the completeness values for individual detections are provided in Table \ref{tab:coeff}. Following our expectations, the width of the emission line significantly impacts the detectability. We will most likely recover broad and strong emission lines, for which the detection fraction reaches $80 - 90$ per cent. For emission lines below 5$\sigma$, we reach a lower level of completeness.

Our quoted completeness values appear lower than those reported in previous surveys \citep[e.g., ASPECS][]{aspecs}. However, notwithstanding the low completeness, two candidates with optical data have confirmed counterparts, which means that these candidates are reliable. 

We define the completeness coefficient as the fraction of retrieved sources for a given peak signal-to-noise and emission-line width (Table \ref{tab:coeff}). Next, the uncertainty on the completeness coefficient measurement is calculated as the 1-$\sigma$ Poisson limit in the signal-to-noise and width bins. Finally, the completeness fraction deduced from this analysis is used to correct the mass density function. 
We assume all our detections are unresolved in the completeness procedure described above. However, especially in the case of low-redshift objects, we can potentially come across marginally resolved sources (of a size 2$-$3 times the synthesized beam). To verify if the same completeness coefficient applies to marginally resolved sources,  we repeated the procedure described above for mock sources with spatial sizes  2, 3, or 4 times the beam size. We find that regardless of the size of the source, the completeness coefficients are within the error. Therefore we are confident that the marginally resolved sources of $S/N$  > 4 are well represented in our search.

\subsection{Reliability}
\label{sec:reliability}
\begin{figure}

	\includegraphics[width=\columnwidth]{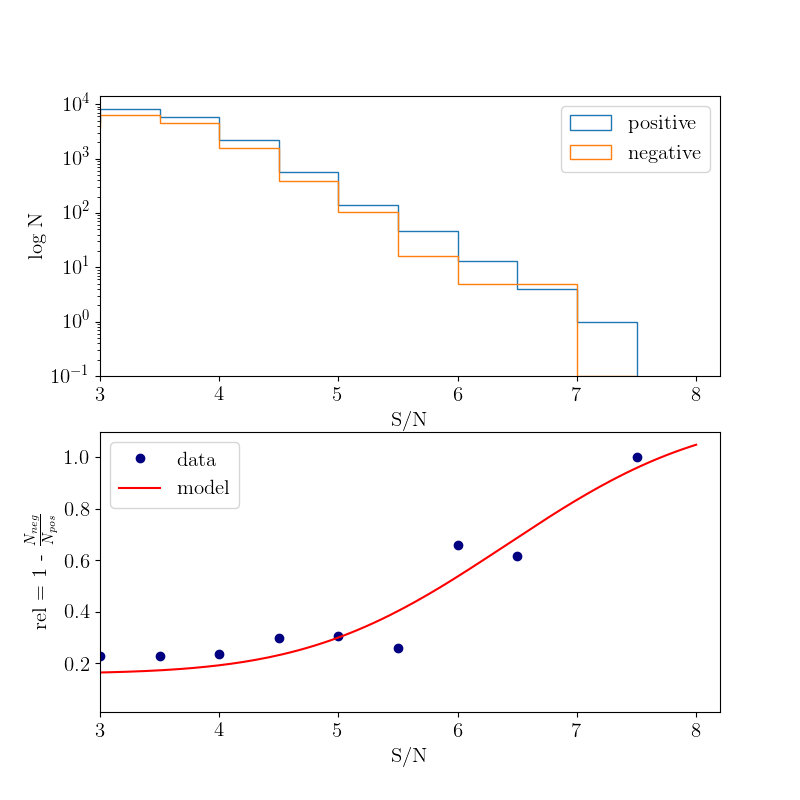}
    \caption{Calculating the reliability coefficient. The top panel shows the distributions of peak signal-to-noise for the positive and negative detections. The surplus of positive detections defines the reliability coefficient. The bottom panel shows the reliability model (Eq.\ref{eq:model}) fitted to the ratio of negative versus positive detections (Eq.\ref{eq:ratio}). Thanks to the model, we can calculate the reliability coefficient at any signal-to-noise ratio.} 
    \label{fig:reliability}
\end{figure}

Assessing the significance of the detection is one of the challenges of any untargeted survey. Especially when using interferometric data, disentangling noise peaks from the real detections becomes a challenge (with the non-gaussian characteristic of the ALMA noise). The most common practice of assigning a fidelity or reliability parameter (as they are interchangeably named) comes from analyzing the number of positive and negative detections in the data cubes. The negative detections are searched for in the inverted cube. As we do not expect real absorption features randomly in the field, these detections represent the distribution of the noise peaks. By comparing several positive and negative detections as a function of their signal-to-noise ratio (Fig.\ref{fig:reliability}), we construct the reliability coefficient, stating which percentage of the detections are likely real.
In our survey, we adopted an approach similar to the method employed by the ASPECS survey, presented in \citet{walter16}. In the ASPECS survey, the reliability threshold was chosen at 60 per cent. In later works, ASPECS updated their method, by including the line width for the reliability calculations of the candidates. However we do not have a large enough sample to adopt this approach, and settled on a simpler method described below.

We define the reliability coefficient as 

\begin{equation}
\label{eq:ratio}
 {\rm reliability}  =  R(S/N) = 1 - \frac{ N_{\rm  neg}}{ N_{\rm pos}},
\end{equation}

where $N_{\rm pos}$ and $N_{\rm neg}$ are the number of the positive and negative candidates for a certain $S/N$.  
To make the reliability calculation more general, we fit the model error function to the distribution of such calculated reliability with $S/N$:
\begin{equation}
\label{eq:model}
     R(S/N) = \frac{1}{2} {\rm erf} \bigg ( \frac{S/N - C }{\sigma} \bigg ) + a,
\end{equation}
which allows us to calculate the reliability parameter for any candidate.  
The model fit to the data (Fig.~\ref{fig:reliability}) results in the model parameters: $C = 6.24$, $\sigma = 2.56$ and $a = 0.65$.
In this approach, we reach a reliability of 0.6 at $S/N = 6.1$ and 0.8 at $S/N  = 6.9$. We require candidates to have a reliability of at least 0.3 to be included in our analysis.

\begin{table*}
	\centering
	\caption{ALMACAL-CO emission-line candidates. The two detections in the field of calibrator J0334-4008 coincide with nearby galaxies detected with VLT/MUSE, allowing for the redshift identification of the sources as CO(1$-$0) lines. For these two detections we assign the probability function of 100 per cent for the spectroscopically confirmed transition.  The columns: (1) Candidate ID including sky coordinates of the detection in J2000, (2) central frequency of the detection in GHz, (3) integrated line flux, (4) width of the detection, (5) signal-to-noise as reported by the SoFiA source finder, (6) Primary Beam attenuation at the position of the detection, (7) distance from the phase centre, (8) Continuum flux limit (index indicating the ALMA beam in which the continuum was measured). }
	\label{tab:candidates}
		\renewcommand{\arraystretch}{1.2}
	\begin{tabular}{lccccccc} 
		\hline \hline
		Source &  Frequency & Flux  & FWHM  & S/N & PB  & $\Delta r$ & Cont.  \\
         &   [GHz] &  [mJy km s$^{\rm-1}$] & [km s$^{\rm-1}$] & & attenuation  & [arcsec] & flux [mJy] \\
		\hline

ALMACAL J000614.36$-$062329.4 &	348.46 &	50 $\pm$ 10 &	35 $\pm$ 5 & 5.9 & 0.45 & 9.10 & < 0.08$_{B7}$\\
ALMACAL J051002.35$+$180052.5&	344.57 &	80 $\pm$ 13	& 35 $\pm$ 4  & 5.1 & 0.30 & 11.13 &< 0.08$_{B6}$ \\
ALMACAL J192450.65$-$291444.7 &	233.48 & 40 $\pm$ 13 &	60 $\pm$  15 & 4.9 & 0.36 & 15.66 &< 0.06$_{B6}$ \\
ALMACAL J235752.46$-$531120.7 &	135.91 &	60 $\pm$ 10	& 200 $\pm$ 26 & 6.7 & 0.64 & 18.20 & < 0.03$_{B4}$ \\
\hline
ALMACAL J033416.50$-$400816.0  &	97.23 &	    100 $\pm$ 22	& 255 $\pm$ 43 & 5.9 & 0.40 & 33.55 & < 0.05$_{B3}$ \\
ALMACAL J033412.22$-$400806.8 &	101.67 &	40 $\pm$ 7	& 60 $\pm$  9 & 6.4 & 0.62 & 25.32 & < 0.06$_{B3}$ \\   
\hline
\hline
	\end{tabular}
\end{table*}

\section{Estimation of CO candidate redshifts}
\label{sec:redshift_prob}

The challenge of using ALMACAL data for an untargeted CO emission-line search is the uneven spectral coverage of each field. All the detected candidates are single-line detections, creating a degeneracy between redshift estimation and transition classification. Due to the limited spectral coverage per field, the probability of detecting more than one transition from the same objects is low.

As shown in Figure~\ref{fig:j0334}, we unambiguously determined the redshift of two of our detections. To classify our other candidates, we employ a novel probabilistic approach based on the catalogue of properties of simulated galaxies with the newly developed open-source, highly complex detailed physics code Semi-Analytical Model (SAM) \textsc{Shark} \citep{lagos18,lagos19}. 

\subsection{Redshift probability function based on \textsc{Shark} SAMs}
We used \textsc{Shark} Semi-Analytical Models (detailed description of the simulation in Section \ref{sec:Shark}) of galaxy evolution to create the \textit{redshift probability calculator}, returning a redshift probability function for each detection, describing the likelihood for the candidate to be of a given CO transition. 

The CO excitation models used by \textsc{Shark} for the brightness predictions come from \citet{lagos12}. These CO simulations combine the galaxy formation model GALFORM with the Photon Dominated Region code UCL-PDR in the $\Lambda$CDM framework (from $z = 0$ to $z = 6$). The H$_2$ and HI gas content of galaxies is predicted by GALFORM \citep{lagos11a, lagos11b} based on the semi-empirical relation from \citet{blitz06}. Combined with the simulated galaxies' ISM properties, these gas masses are further converted to the CO emission (transitions from $1-0$ to $10-9$) through UCL-PDR \citep{bayet11} models for each galaxy. The resulting CO Spectral Lines Energy Distributions or SLEDs depend strongly on the galaxy ISM properties and scale with the galaxy's IR-luminosity.

We generated a catalogue of \textsc{Shark} simulated galaxies, requiring the brightest CO transition to be brighter than 0.05 mJy km s$^{-1}$. The selection resulted in a catalogue of about 10 million sources covering redshifts from $z = 0 - 6$. For each CO transition, we created a 2D histogram binning the data of CO transition flux and redshift. For each bin of transitions, $J_{\rm up} > 1$, we additionally save the CO(1$-$0) flux corresponding to each bin. This way, we use the CO SLEDs from the simulated galaxy and do not require empirical line ratios. Lastly, we divide the number of objects in each bin by the total number of objects in the simulated sample defining the probability coefficient for a certain pair of flux and redshift. 

The CO(1$-$0), the transition whose luminosity is converted directly to molecular gas mass, is observable with ALMA up to redshift $z = 0.35$, and due to the small probed volume, we do not expect to detect many galaxies from that line. On the other hand, high-level transitions ($J_{\rm up}$=8,9,10) are expected to be faint for the typical star-forming galaxy ISM conditions, and their probability is close to or exactly zero. From the \textsc{Shark} probability functions, the most likely transitions corresponding to the detections are $J_{\rm up}$ = 3,4,5 and 6, which places most of our candidates between redshift 0.5 and 1.5. For each of the probable redshift predictions, we check its position with respect to the calibrator redshift (which could raise concerns about the clustering). We found that none of the potential redshift classifications is close to one of the corresponding calibrators (see Fig. \ref{fig:det}). Therefore, in our classification of the detections, we do not adopt the highest probability $J$-transition as the final classification. Instead, we include all probable $J$-transitions, weighted by the probability predicted by \textsc{Shark}, in our calculations. 

Our approach is not very sensitive to the details of the models. However, the essential requirements of the model are that it is in qualitative agreement with the observed H$_2$ content of galaxies across cosmic time \citep[shown to be the case in][]{lagache18}, and with the observed CO SLEDs (shown in Appendix \ref{sec:Shark_sled}).

\subsection{CO SLEDs of \textsc{Shark} galaxies}
\label{sec:Shark_sled}
\begin{figure}
	\includegraphics[width=\columnwidth]{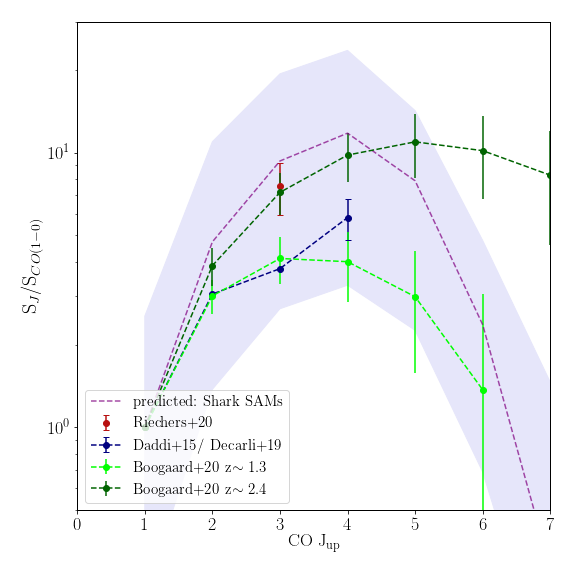}
    \caption{The average CO SLED of BzK $\langle z \rangle$ = 1.5 galaxies simulated by \textsc{Shark}. The dashed purple line marks the mean SLED of the subset chosen to this test, the lavender area covers the 16th and 84th percentiles. The dark blue points and dashed line are a CO SLED averaged over four BzK galaxies at $\langle z \rangle$ = 1.5 from \citet{daddi15}. The red point shows the median measurement form \citet{vlaspec}, while the green show the modelled SLEDs for galaxies at $<z> = 1.3$ and $<z> = 2.4$ from \citet{boogaard2020}. The fluxes of all transitions are normalised to the CO(1$-$0) flux. 
   }
    \label{fig:sled}
\end{figure}

The redshift estimates of the ALMACAL-CO detections are based on the CO emission-line flux predictions from the \textsc{Shark} simulations \citep{lagos18, lagos19, lagos20}. The relative strength of the CO lines in the CO ladder is described by the CO SLED (Spectral Line Energy Density) and depends on the properties of the host galaxies, especially their ISM conditions. \citet{lagos20} compared the predicted CO SLED of SMG and regular main-sequence galaxies to a compilation of observations, finding broad agreement. Thus, we expect that using SHARK CO SLEDs here would provide a broadly correct framework to estimate possible redshifts of our detected sources.

We want to test \textsc{Shark} CO SLEDs predictions against observations to see how well the observable galaxies are represented. The SLEDs directly impact the results of our survey as they determine the relative brightness of the high-$J$ lines with respect to the CO(1$-$0). First, we base our redshift probability estimates on the fluxes of the emission lines as we compare the brightness of the observed line with simulated line fluxes of different CO transitions. Secondly, we converted observed CO luminosities to CO(1$-$0) transitions (used for the molecular gas mass estimation) using the simulated values, which are also dependent on the CO SLEDs predicted by simulations. 

We know of only a handful of CO SLEDs in star-forming galaxies and some extreme objects like quasars, SMGs, or starburst galaxies. So far, the most extensive study is that of CO SLEDs from BzK galaxies at a median redshift of $\langle z \rangle$ = 1.5 in \citet{daddi10}. The authors report the flux measurements for four CO transitions ($J_{\rm up}$ = 1,2,3,4) in four galaxies creating an average CO SLED, which we used for the comparison with the SLEDs predicted by \textsc{Shark}. BzK galaxies represent a typical star-forming object in the Universe above $z >$ 1.0. Since most of our detections are associated with galaxies at $z= 1-2$, we consider them a representative comparison sample for our results.

Following the optical selection criteria of BzK galaxies from \citet{daddi10}, we extracted a set of BzK galaxies from the \textsc{Shark} catalogue and calculated their median CO SLED together with their 16th and 84th percentiles. The resulting SLED is presented in Fig. \ref{fig:sled}. We also plotted the observational results from \citet{daddi15}. The \textsc{Shark} simulated SLEDs are in good agreement with the observational points. Both SLEDs peak at the $J_{\rm up}$ = 4, and the observed points lie within the envelope of simulated values. 

Recently, new VLA observations from \citet{vlaspec} have shown that the \citet{daddi15} SLEDs tend to overestimate the CO(1$-$0) luminosity and consequently have overestimated molecular gas masses of galaxies at redshift $z \sim 2-3$. Additionally \citet{boogaard2020} shown that the slope of the CO SLEDs vary significantly beetween galaxies at redshfits $z \sim 1$ and $z \sim 2$. In Figure \ref{fig:sled}, we compare the median of these measurements to \textsc{Shark} SLED and the \citet{daddi15, vlaspec, boogaard2020} results. This updated measurement is also in agreement with simulated SLEDs. The results show that the simulated SLEDs in this work reproduce well the physical parameters of CO-selected galaxies. 

For this probabilistic approach, one could ideally consider using a database of observed CO emission lines in red-shifted galaxies. However, currently available observations of CO emission (especially multi-CO line observations from the same object) are not extensive enough to realize such an approach. We conclude that the SAM approach is currently the only viable option and have demonstrated that applying this approach does not lead to any significant biases.

\subsection{Lowest possible $J$-transition limit}

The main measurements presented in this work rely on the simulated CO SLEDs obtained from the \textsc{Shark} simulations. Additionally, to constrain limits on our simulation-based transition classification, we explore a conservative \textit{lowest-$J$} assumption. In this approach, we assign the detection with the lowest possible $J_{\rm CO}$ transition observable at the observed frequency. We expect the transitions below $J_{\rm up}= 5$ to be the brightest and most probable ones to be observed. The transitions assigned in this way range from $J_{\rm up}$=1 to $J_{\rm up}$=4.

Assuming the lowest $J$ transition places all detections at the lowest possible redshift. That places five out of six detections below redshift $z <0.5$ and one to the next redshift bin ($z = 0.5-1.0$). We bin all the lowest $z$ detections and include them as the \textit{lowest-J} estimate of the $\rho(M_{\rm H_2})$ (Fig. \ref{fig:mh2}).

To calculate the luminosity of the CO(1$-$0) transition and corresponding $M_{\rm H_2}$, we need to assume a CO-SLED. Only a limited number of data-based CO-SLEDs in the literature cover the low-$J$ transitions, including the CO(1$-$0), which can be used for molecular gas mass estimates and here we adopted a median CO-SLED from \citet{daddi15}. The usual choice of \citet{daddi15, vlaspec} SLEDs, used by ASPECS, are derived for BzK galaxies at redshift $z = 1.5$ and are not suitable in this case. We apply the SLED transitions coefficients for the \textit{lowest-J} limit. 

This result provides a conservative model-free limit to complement the simulation-based line classification. We emphasize that the \textit{lowest-J} approach explored here should not be considered a measurement but should only be seen as a conservative limit to explore the maximum effect of unknown $J$-transitions.

\subsection{Search for optical counterparts}

Based on the \textsc{Shark} Semi-Analytical Models, the probabilistic approach provides population-weighted predictions for the observed lines. However, identifying optical counterparts would be desirable for a fully secure classification of each CO detection. Therefore, we searched several optical data archives for observations of the calibrator fields. 

In general, we expect the optical counterparts of the CO line detections to be similar to ASPECS counterparts \citep{boogaard}, rather faint ($\sim$ 23--25~mag) sources. We checked the archives of the optical all-sky Pan-STARSS1 \citep{panstarss} survey for the presence of any continuum detection at the position of ALMACAL-CO line candidates. We did not detect any optical source at any candidate's position. With exposure times of the order of a few minutes, Pan-STARRS1 observations are likely not deep enough to detect the counterparts of our detections. In near-infrared data from VISTA, which are available for four detections, we also do not detect counterparts for our detections down to the level of 20.6-21.7$^m$ in $J$ or to 20.4$^m$ in the $Ks$ filter. For two fields (J0510+1800 and J1651+0129), only ALMA data are available.

As we have already shown in Fig.~\ref{fig:j0334}, archival MUSE data are available for one of our fields named J0334-4008, in which the two ALMACAL detections coincide spatially with two bright galaxies at redshifts $z_1 = 0.133$ and  $z_2 = 0.185$, securely placing the line candidates as  CO(1$-$0) transition at these redshifts. Those systems have confirmed redshifts, and for further analysis, we adopt that redshift solution with 100 per cent probability. We note that in the accompanying redshift probability analysis, the solution with $J_{\rm up}$ = 1 is indeed assigned a significant probability (the highest probability is CO(2$-$1) at $z \sim$ 1.2). Detecting the optical counterparts to these sources confirms the robustness of the CO emission lines reported in this study. 

Additionally, we searched the ALMACAL archives for the other transitions corresponding to the candidates, however, we do not report any detections (full discussion in Appendix \ref{sec:app_cand}).






\begin{figure*}
	\includegraphics[width=2\columnwidth]{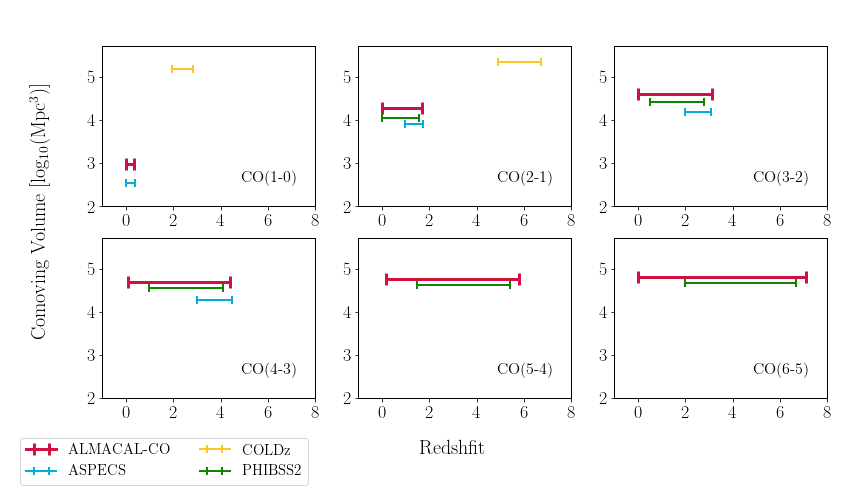}
    \caption{Comparison of volume coverage for CO transitions from $J_{\rm up}=1$ to $J_{\rm up}=6$ in ALMACAL-CO pilot (red), ASPECS - blue \citealt{aspecs}, PHIBSS2 - green, \citealt{phibbs2} and COLDz - yellow \citealt{coldz}). ALMACAL-CO pilot covers a larger volume for individual transitions (except for high redshift COLD-Z) thanks to the large frequency span.}
    \label{fig:volume}
\end{figure*}


\section{Molecular gas mass density estimate}
\label{sec:volumes}

To derive a reliable measure of the evolution of the cosmic molecular gas mass density $\rho(M_{\rm H2})$, we first need to evaluate the co-moving volume probed by our survey. This volume needs to be calculated as a function of redshift and CO luminosity separately for each CO transition probed by the survey. It can be achieved by determining the co-moving volume elements $dV$, corresponding to redshift element $dz$ and solid angle element $d\Omega$:

\begin{equation}
    dV_c = D_H \frac{(1+z)^2D_A^2}{E(z)}d\Omega dz,
\end{equation}
where $D_H$ is the Hubble distance ($D_H$ = 3000 $h^{-1}$ Mpc), $D_A$ is angular diameter distance at redshift $z$ and $E(z)$ is the scaling factor defined as: 
\begin{equation}
E(z) = \sqrt{\Omega_{\rm M}(1+z)^3 + \Omega_{\rm K}(1+z)^2+ \Omega_{\Lambda}}.
\end{equation}

The CO luminosity defining the limits of the volume integration depends on the sensitivity, which is not uniform across the field of view but instead drops off gradually from the centre of the primary beam, and can be modelled with a Gaussian function. Therefore the total cosmic volume accessible to a hypothetical galaxy of a certain luminosity would have a cone shape. To estimate the lowest detectable luminosity, we use a modelled $v$ = 200 km s$^{\rm -1}$ boxcar shape emission line with a maximum flux of 5 $\times$ the r.m.s of the cube. We convert these parameters to an $L'_{\rm line}$ luminosity following \citet{carilli} and calculate the volumes for for a range of line luminosities from 10$^{5}$ to 10$^{13}$ K kms$^{-1}$pc${^2}$. 

In addition to the sensitivity, for some CO transitions volume might be limited by the cube's frequency coverage (which translates to redshift) instead of sensitivity. In this case we integrate volume elements by summing the contributions of $dz$ - thick rings of increasing radius defined by the observed frequency. 

Finally, we sum volumes calculated for each data cube to determine the total survey volume as a function of luminosity. These comoving volumes are determined for each CO transition from CO(1$-$0) to CO(6$-$5), summing the volume coverage of each of the cubes for redshift from 0 to 5.

The results are presented in Table \ref{tab:volumes}. ALMACAL-CO sample is dominated by the high-frequency datacubes (Band 6, 7, 8) corresponding to large volumes probed by the high-$J_{\rm CO}$ transitions. The frequency coverage is complementary to the work of previous untargeted surveys \citep{aspecs,coldz,phibbs2}, mapping the different CO transitions and redshift parameter space (see comparison in Figure \ref{fig:volume}). ALMACAL-CO probes comparable volumes of $J_{\rm up}$ < 4 transitions, typically targeted by other surveys. Thanks to the extensive frequency coverage, we reach a 2$-$5 times larger total volume than previous works.

\begin{table}
	\centering
	\caption{Volume per CO transition covered by the ALMACAL-CO sample. Depending on the frequency coverage each transition is traced in different redshift bin. The dependence of the volume on the CO luminosity is shown in Fig. \ref{fig:volume}. Columns: (1) covered CO transitions, (2) - redshift coverage of the transition, (3) Co-moving volume corresponding to the transition.  }
	\label{tab:volumes}
	\begin{tabular}{ccc} 
		\hline \hline
		Transition & Redshift & Volume \\
 &  & [cMpc$^{3}$] \\
		\hline
CO(1$-$0) & 0.00--0.35 & 958 \\
CO(2$-$1) & 0.00--1.70 & 18746 \\
CO(3$-$2) & 0.00--3.15 & 38675 \\
CO(4$-$3) & 0.10--4.40 & 49817 \\
CO(5$-$4) & 0.20--5.80 & 58639 \\
CO(6$-$5) & 0.01--7.10 & 64742 \\
		\hline
total & 0.00--7.10 &  231577 \\
\hline  \hline
	\end{tabular}
\end{table}

\begin{figure*}

	\includegraphics[width=2\columnwidth]{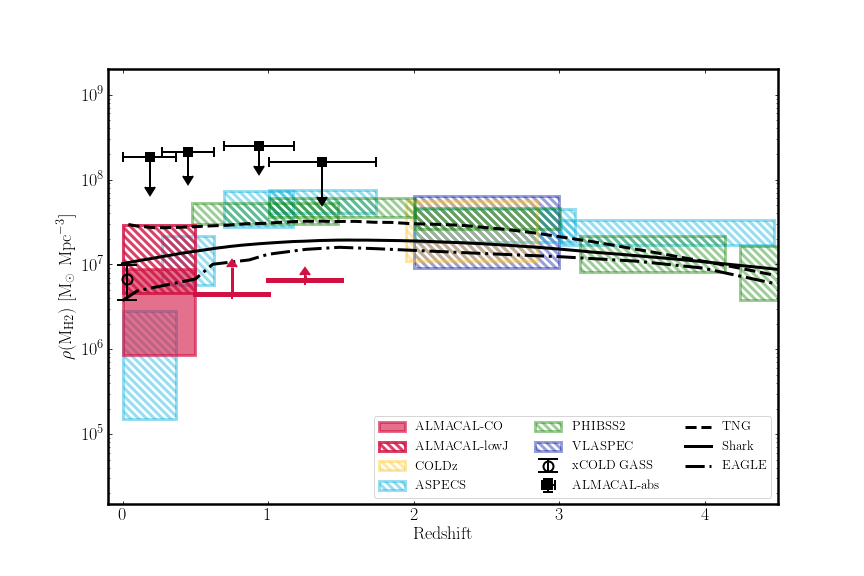}
    \caption{The evolution of molecular gas mass density with redshift as measured with ALMACAL-CO pilot survey (red boxes). The filled box represents the results of the \textsc{Shark}-based source classification, while the red dashed box corresponds to the \textit{lowest-J} approach. We present limits for the highest redshift bins due to the insufficient number of detections populating these bins. The results of other surveys are also shown: ASPECS \citep[][in blue]{decarli20} COLDz \citep[][in yellow]{coldz}, PHIBSS2 \citep[][in green]{phibbs2}, and VLASPEC \citep[][in dark blue]{vlaspec}. The empty black circle marks the local measurements of molecular mass content of galaxies from xCOLD GASS \citep{fletcher}. All measurements (including ALMACAL-CO) are shown with the 1-$\sigma$ Poissonian uncertainty. Black squares are the limits from a molecular absorption-line search in ALMACAL by \citet{klitsch19} (marked as ALMACAL-abs in the legend). Black lines mark the predictions of evolution of  $\rho(M_{\rm H2})$ from simulations: dashed line - IllustrisTNG \citep{popping19}, dashed-dotted line - EAGLE \citep{lagos15}, solid line - \textsc{Shark} \citep{lagos18}. The simulated evolution represents the results of the ALMACAL-CO pilot, especially the low redshift bin. Our results are consistent with the findings of the previous surveys and simulation predictions and show the power of the new simulation-based approach to classify the sources lacking the ancillary data.}
    \label{fig:mh2}
\end{figure*}

\begin{table}
	\centering
	\caption{ALMACAL-CO survey measurements of the $\rho(M_{\rm H2})$. The presented range represent 1 $\sigma$ confidence boundaries. We do not include the poorly populated redshift bins in the calculations. In the last row we present the result of the \textit{lowest-J} approach.
	 Columns: (1) redshift bin, (2)   $\rho(M_{\rm H2})$ 3$\sigma$ range calculated for that redshift bin. }
	\label{tab:omega}
	\begin{tabular}{ccc} 
		\hline \hline
		Method & Redshift bin &   $\rho(M_{\rm H2})$ \\
		& & $1 \times 10^{7} \rm M_{\odot}$ \\
		\hline
\textsc{Shark} & 0.0--0.5 & 0.08--0.87 \\
&0.5--1.0 & 0.98--3.24 \\
&1.0--1.5 & 0.68--2.63 \\
\hline
\textit{{lowest}-J} & 0.0--0.5 & 0.24--2.75 \\

\hline \hline
	\end{tabular}
\end{table}

\subsection{The molecular gas mass density}
\label{sec:colum}

The ultimate goal of ALMACAL-CO is the measurement of the evolution of molecular gas mass density with redshift. Once we have analysed the entire dataset (in a forthcoming publication), we aim to constrain the CO luminosity function with high accuracy at different redshifts. In the current pilot project, we introduce a novel statistical approach relying on the probabilistic redshift association of our sources (see Section \ref{sec:redshift_prob}). Since every detection is assigned a redshift probability function, we include all possible CO transitions in the molecular gas mass density calculations by weighting the CO luminosity by its derived probability. We stress that our survey results at the current stage are preliminary, and we cannot reconstruct the full CO(1$-$0) luminosity functions (LFs) with the existing data. 

To construct the CO(1$-$0) space densities for the detections reported here, we calculate 1000 realisations of each probability function, constructing the final statistical sample for our studies. We divide the detections into three redshift bins (resembling the lowest redshifts bins used in the ASPECS survey): $\langle z_1 \rangle$=0.25, $\langle z_2 \rangle$=0.75, $\langle z_3 \rangle$=1.25.  

The majority of our detections are classified as relatively high-$J$ transitions. To unify the luminosity function, we use the CO(1$-$0) brightness predicted by the \textsc{Shark} averaged CO SLEDs corresponding to each detection, which we then convert to line luminosity using the assigned redshift and simulated signal integrated flux \citep{carilli}. Note that assuming observed CO SLEDs of normal star-forming galaxies would yield the same results, as \textsc{Shark} reproduces very well the observed CO SLED of normal star-forming galaxies selected from the colour-colour method of \citet{daddi15} as well as recent measurements from VLASPEC \citep{vlaspec}. Finally, we discuss the details of the simulated CO SLEDs and the comparison to observations in Section \ref{sec:Shark_sled}.

Molecular gas masses are obtained through scaling the CO(1$-$0) luminosity with the $\alpha_{\rm CO}$ conversion factor. In this work we adopt the Milky Way CO conversion factor $\alpha_{\rm CO}$ = 4.3 M$_{\odot}$/(K kms$^{-1}$ pc$^{2}$) for all detections \citep{alphaco}. 


For each redshift bin, we can now construct the molecular gas mass density $\rho(M_{H_2})$:
\begin{equation}
\label{eq:mass}
    \rho(M_{H_2}) = \alpha_{\rm CO} \sum_{i=0}^{N} \frac{R_i}{c_i} \frac{L_{i'_{\rm CO(1-0)}}}{V_i},
\end{equation}
where 
$R_i$ is the reliability of the detection (Sec.\ref{sec:reliability}), $c_i$ is the completeness factor evaluated in \ref{sec:completeness}, $V_i$ is the volume accessible to each detection, and $i$ is the iteration over the detections in the redshift bin. 

We repeat the calculations over all 1000 projections of the probability function of each candidate and eventually we take the mean value for each luminosity bin. 
We estimate the statistics uncertainty in each redshift bin following  Poissonian low number statistics \citep{Gehrels86}. 

In the case of the \textit{lowest-J} assumption, we apply the \citet{daddi15} CO SEDs to calculate the CO(1$-$0) flux corresponding to the assigned transition. Following Equation \ref{eq:mass} we calculate the CO(1$-$0) molecular gas mass density. As five of the six detections have a corresponding redshift below $z = 0.5$, we bin them in one redshift bin. We calculate the limits assuming all six candidates have the lowest (lower limit) and the highest (upper limit) luminosity. Binning all of the detections in the lowest redshift bin results in a higher $\rho(M_{\rm H2})$ estimate than from the \textsc{Shark}-based approach (Fig. \ref{fig:mh2}) and measurements from the literature. 

In Figure \ref{fig:mh2}, we include only molecular gas mass density estimates based on the untargeted CO observations (both in emission and absorption). There are references in the literature providing the estimates on the molecular gas mass density based on a different approach, which we did not include on the plot, e.g., scaled dust continuum, based on empirical gas-to-dust conversion \citep{scoville, liu19,magnelli,garratt2021, Wang2022}, radio continuum to CO luminosity conversion \citep{orellana} or CO intensity mapping \citep[see discussion in ][]{poppingCP}. 

\section{Discussion and Conclusions}

The ALMACAL-CO pilot project described in this work is a proof-of-concept study, introducing a different approach to the sub-mm untargeted CO emission-line survey. We report six emission line detections found in the 33 calibrator fields, each with an accumulation integration time exceeding 40 minutes.

A characteristic of our survey is that our fields are centred on bright sub-mm ALMA calibrators and are not cosmological fields observed in a wide range of wavelengths, therefore, we lack ancillary data to classify the detections. Anticipating that this problem will hold for the full ALMACAL-CO survey, we also tested a novel statistical approach based on the probabilistic predictions from Semi-Analytical Models. In this way, we do not focus on the precise classification of each source, instead, we adopt a redshift probability distribution function for each detection, and through the statistical evaluation, we arrive at the result averaged over the realisation of the PDF. 

We adopt a probabilistic approach by constructing a redshift probability function for each candidate complemented by frequency-coverage-based lowest possible $J$-transition classification. Based on the predictions from the \textsc{Shark} SAMs, we assign corresponding CO transition probabilities to each of the detections. We run a set of tests to assess how these predictions compare with observations and show that the SAMs predictions align with results from previous observational surveys. To provide limits to the simulation-based approach above and test the uncertainties, we also evaluate the results assuming the lowest possible $J_{ \rm CO}$ observable in the frequency range given by the data for each candidate.

We run a set of tests to assess how the predictions from \textsc{Shark} SAMs compare with the observations and if the simulated CO SLEDs are comparable with ones observed in previous surveys. We found that the SLEDs predicted by \textsc{Shark}, which affect directly our redshift probabilities, are comparable to the ones observed in star-forming galaxies (see Appendix \ref{sec:Shark_sled} for comparison with \citealt{daddi15} SLEDS of BzK galaxies). We also run a test on the subset of ASPECS detections with confirmed redshifts \citep{aravena19} and PHIBSS2 objects with known optical counterparts \citep{phibbs2} through our redshift probability calculator (details of the test can be found in Appendix \ref{sec:test_Shark}). The probability functions assigned to detections of these surveys predict a correct transition for 40 per cent ASPECS and 60 per cent for PHIBSS2 detections. The results of these tests support the reliability of the SAMs-based approach.

As calibrator data are not as homogeneous as the science program observations, the noise properties and uneven frequency coverage make the survey more challenging in some aspects. Nevertheless, in the subset of the deepest observations, we detected six candidates making a case for an ALMACAL-wide search, however, the untargeted observations at redshift $z<1$ suffer because only small cosmological volumes are probed. Hence, they might not represent the Universe and require further observational studies. In addition, more stringent constraints are required to determine the shape of the molecular gas mass density curve.

The survey covers a frequency range higher than that used in previous CO surveys such as ASPECS, COLDz, and PHIBSS2. By probing ALMA Bands 6, 7, and 8 (211$-$500~GHz), we are more likely to observe high-$J$ transitions, which are expected to be fainter than the CO(2$-$1) and CO(3$-$2) lines typically found in Band 3 (see Fig. \ref{fig:det} for probability functions corresponding to each detection). Thanks to the broad frequency coverage (from Band 3, 84~GHz to Band 8, 500~GHz), this study probes several CO lines, from $J_{\rm up}$ = 1 to $J_{\rm up}$ = 6 (and above, however, due to the low brightness of the high-$J$ transitions we do not expect to be sensitive to those). In principle, the ALMACAL-CO pilot survey probes a volume larger than that of previous projects by a factor of two,  cumulatively over all fields, the survey also probes a larger area. 

As a result of our probing higher-$J$ transitions, our detections are both fainter and have narrower profiles than the typical sources observed by other surveys (Fig. \ref{fig:flux}). We are probing the limits of the detection reliability for sub-mm interferometric data and a fraction of the presented detections are likely bright noise peaks. We attempt to minimize this by carefully inspecting all secured detections and remove about half of those as their channel maps were highly suggestive of being spurious in nature. At this stage of the project, we are unable to assess the real reliability of these detections further, due to the lack of secondary lines and optical counterparts. We discuss in detail the efforts to understand the nature of our detections in Appendix \ref{sec:app_cand}. On the other hand, one of our narrowest detections (ALMACAL J033412.22-400806.8) is confirmed with HST and MUSE observations to be associated with a face-on spiral galaxy, proving that the FWHM of the detection alone cannot be a criterion against its reliability (Fig. \ref{fig:j0334}). 

The cosmic molecular gas mass density derived from our ALMACAL-CO survey follow the trends indicated by the previous untargeted surveys. Our constraints on the measurements below $z < 1.5$ agree with the literature. For example, in the lowest redshift bin, we agree with the measurements of \citet{fletcher} and the upper limits from the CO absorption search from \citet{klitsch19}. On the other hand, at $z = 0.5$, ALMACAL-CO measures a lower molecular gas mass density than previous observations.  Due to the low number of candidates which line classification places them at the redshift $z > 1.5$ we cannot provide constraining measurements of  $\rho(M_{\rm H2})$ at this redshift range. The lower limits are consistent with the measurements from the literature and the simulations. A combination of factors contributes to the low number of sources detected at these redshifts: a small field-of-view at high frequencies, expected low fluxes for the high-$J$ transitions detectable at the probed frequencies and an insufficient depth of the observations. 

The evolution of $\rho(M_{\rm H2})$ predicted by hydrodynamical simulations such as EAGLE and IllustrisTNG \citep{lagos15, popping19, maio} is less dramatic in magnitude than observed so far. None of the available simulations reproduce the strong evolution of the molecular gas mass function inferred by ASPECS. Our results agree well with previous observational inferences and confirm the tension above.

The results of ALMACAL-CO, taken together with other surveys, cannot currently distinguish between two opposing scenarios: an evolution of molecular gas with a redshift that follows the evolution of SFH or a weak evolution, similar to that observed for $\Omega_{\rm HI}$. Neutral gas dominates molecular gas at all redshifts by at least a factor of 2--3 \citep{peroux20, szakacs22}. However, the molecular gas mass density does not evolve as dramatically as the SFH. 

In the future, we aim to confirm the ALMACAL-CO pilot detections and constrain their redshifts by searching for optical/IR counterparts and conducting targeted ALMA follow-up of the remaining unconfirmed detections. After the success of the pilot study, we will expand the untargeted search to all calibrator fields, including all new data collected since the original selection of the pilot sample. This will expand the area of the future survey to $\sim$600 arcmin$^{2}$ and increase the probed volume $\sim$50 times. With the future extension to the total ALMACAL dataset, we will improve the statistics and expect significantly more line detections, resulting in better constraints of the $\Omega_{\rm H_2}$. With this significant survey expansion, we expect to detect 50--100 CO emitters. To complement the process of collecting the ancillary data of the CO-selected galaxies, we will use the Semi-Analytical Models (SAMs) to place constraints on the detection identification and use them as predictions for the follow-up observations. 

\section*{Acknowledgements}
The authors thank I.R. Smail for his contribution to the analysis and discussion as well as  T. Westmeier, G. Popping, and R. Decarli for useful discussions and comments. AK gratefully acknowledges support from the Independent Research Fund Denmark via grant number DFF 8021-00130. This work was supported by resources provided by the Pawsey Supercomputing Centre with funding from the Australian Government and the Government of Western Australia. This paper makes use of the following ALMA data: ADS/JAO.ALMA\#2012.1.00077.S, 2012.1.00105.S, 2012.1.00129.S, 2012.1.00173.S, 2012.1.00275.S, 2012.1.00346.S, 2012.1.00385.S, 2012.1.00621.S, 2012.1.01123.S, 2013.1.00116.S, 2013.1.00139.S, 2013.1.00162.S, 2013.1.00332.S, 2013.1.00359.S, 2013.1.00376.S, 2013.1.00659.S, 2013.1.00726.S, 2013.1.00745.S, 2013.1.00764.S, 2013.1.01010.S, 2013.1.01016.S, 2013.1.01258.S, 2013.1.01271.S, 2013.1.01369.S, 2015.1.00016.S, 2015.1.00030.S, 2015.1.00084.S, 2015.1.00118.S, 2015.1.00122.S, 2015.1.00169.S, 2015.1.00212.S, 2015.1.00274.S, 2015.1.00425.S, 2015.1.00428.S, 2015.1.00480.S, 2015.1.00496.S, 2015.1.00534.S, 2015.1.00607.S, 2015.1.00631.S, 2015.1.00686.S, 2015.1.00702.S, 2015.1.00752.S, 2015.1.00768.S, 2015.1.00964.S, 2015.1.00966.S, 2015.1.00979.S, 2015.1.00981.S, 2015.1.00986.S, 2015.1.01018.S, 2015.1.01020.S, 2015.1.01170.S, 2015.1.01246.S, 2015.1.01268.S, 2015.1.01316.S, 2015.1.01349.S, 2015.1.01406.S, 2015.1.01446.S, 2015.1.01503.S, 2015.1.01512.S, 2015.1.01541.S, 2015.A.00018.S, 2016.1.00010.S, 2016.1.00048.S, 2016.1.00176.S, 2016.1.00248.S, 2016.1.00309.S, 2016.1.00324.L, 2016.1.00344.S, 2016.1.00366.S, 2016.1.00393.S, 2016.1.00406.S, 2016.1.00426.S, 2016.1.00447.S, 2016.1.00470.S, 2016.1.00481.S, 2016.1.00497.S 2016.1.00515.S, 2016.1.00517.S, 2016.1.00533.S, 2016.1.00543.S, 2016.1.00604.S, 2016.1.00605.S, 2016.1.00710.S, 2016.1.00854.S, 2016.1.00870.S, 2016.1.00878.S, 2016.1.01073.S, 2016.1.01155.S, 2016.1.01186.S, 2016.1.01200.S, 2016.1.01209.S, 2016.1.01236.S, 2016.1.01287.S, 2016.1.01293.S, 2016.1.01313.S, 2016.1.01374.S, 2016.1.01505.S, 2016.1.01601.S, 2017.1.00293.S, 2017.1.00300.S, 2017.1.00480.S, 2017.1.00487.S, 2017.1.00497.S, 2017.1.00503.S, 2017.1.00940.S, 2017.1.01223.S, 2017.1.01666.S.
ALMA is a partnership of ESO (representing its member states), NSF (USA) and NINS (Japan), together with NRC (Canada), MOST and ASIAA (Taiwan), and KASI (Republic of Korea), in cooperation with the Republic of Chile. The Joint ALMA Observatory is operated by ESO, AUI/NRAO and NAOJ. This research made use of Astropy,\footnote{http://www.astropy.org} a community-developed core Python package for Astronomy \citep{astropy:2013, astropy:2018}.

\section*{Data Availability}

The data underlying this article will be shared on reasonable request to the corresponding author. The \textit{Shark} code is hosted on GitHub and publicly available (see \citet{lagos18} for details). 

\bibliographystyle{mnras}
\bibliography{biblio}



\appendix
\section{Comparison of \textsc{Shark} SAMs predictions with observations}
\subsection{The \textsc{Shark} Semi Analytical Model of galaxy formation}
\label{sec:Shark}

\textsc{Shark} is based on an N-body hydrodynamical Dark Matter (DM) only simulation, which halos are populated with galaxies without resolving their inner structure. Because SAMs are relatively inexpensive (at least compared to cosmological hydrodynamical simulations), they can be run on very large DM-only simulations, which can then be used to generate mock lightcones with different survey specifications. For this work we use the relatively large area ($\sim$ 108 deg$^2$), extremely deep lightcone (r-band magnitude limit of $\sim$ 30 mags) presented in \citet{lagos19}. This lightcone contains approximately 80 million galaxies from z=0 to z=6. With these properties, these simulations are ideal tools for large scale sky surveys predictions and cross-match between expected and measured quantities including the general evolution of cosmic densities. 

\textsc{Shark} is based on the \textsc{SURFS} N-body DM-only simulation suite, starting at $z$ = 24 \citep{elahi18} and following the \citet{planck16} $\Lambda$CDM cosmology. The properties of simulated halos are then calculated using the\textsc{VELOCIRAPTOR} \citep{elahi19, velociraptor} halo finder and the merger tree builder the \textsc{TREEFROG} \citep{elahi18}. Based on the parameters of the dark matter halos, their merger and accretion history, \textsc{Shark} evolves galaxies using the physical model described in \citet{lagos18}. The model describes the physical processes governing the evolution of galaxies including molecular-gas based star formation, stellar and Active Galactic Nuclei (AGN) feedback, gas accretion, gas shock heating and cooling, galaxy mergers and environmental effects, disk instabilities, chemical enrichment, photoionisation feedback and black hole growth. For the details of the prescriptions and codes used for the creation and evolution of an object in the simulations see \citet{lagos18}. Combining the models with the spectral energy distribution software tool \textsc{PROSPECT} \citep{robotham20} result in the generation of an SED for each simulated galaxy at all simulated snapshots, spanning from FUV to FIR \citep{lagos19}, including dust attenuation and dust re-emission. All galaxies, additionally to their molecular gas masses, have predicted CO luminosity SLEDs following the photon-dissociation region modelling of \citet{lagos12}. Multiple simulation cubes are rotated (to prevent artificial clustering) stacked and evolved over the broad redshift range creating a significant cosmic volume \citep[see][for details of how these lightcones are build]{chauhan19}.

The SAMSs reproduce a wide range of galaxy properties and populations including metal-poor systems and SMGs and reproduce the UV-to-FIR galaxy luminosity functions, the number counts and redshift distribution of sub-millimetre galaxies. The results of the simulations compared with the observations recreate well the cosmic star formation rate density at different redshifts as well as galaxy stellar mass functions.

\subsection{Testing \textsc{Shark} redshift probability calculator with previous detection}
\label{sec:test_Shark}
\begin{figure}
	\includegraphics[width=\columnwidth]{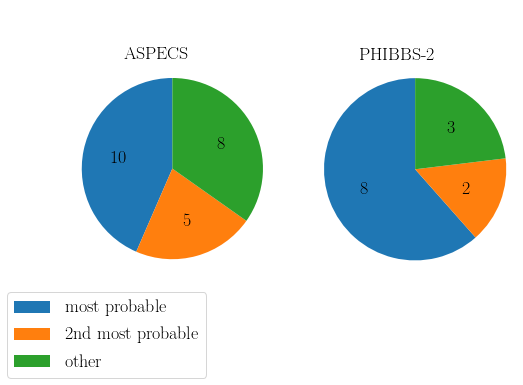}
    \caption{Number of correctly identified transitions of the sources from ASPECS \citep[left,][]{aspecs} and PHIBSS2 \citep[right, ][]{phibbs2} surveys by our \textsc{Shark}-based redshift calculator. The blue slice depicts the number of transitions for which the identification as most probable is consistent with the spectral classification. The orange slice covers transitions for which the most probable $J_{CO}$ was one lower/higher than the classified one (and the actually observed transition was assigned significant probability), while the green slice covers all other cases. The redshift calculator identifies correctly about 50 per cent of the detections (over 60 per cent if we include the "2nd most probable" option).}
    \label{fig:pie}
\end{figure}

Classification of the ALMACAL-CO pilot survey detections relies on the properties of the population generated in \textsc{Shark} Semi-Analytic Models. Since \textsc{Shark} reproduces well the general properties of galaxies in the Universe like galaxy luminosity functions, sub-millimetre number counts and cosmic star formation rate density \citep[see][]{lagos18,lagos19} these predictions are reliable. Before applying the probabilistic approach to ALMACAL-CO candidates we checked the predictions on detections of previous surveys: 23 emission lines from ASPECS \citep{walter16,aravena19} and 13 secondary detections of PHIBSS2 NOEMA survey \citep{phibbs2}. We choose detections with confirmed redshift, through other CO lines or optical counterparts. 

For each of the detections, we construct \textsc{Shark}-based redshift probability function considering all CO transitions. We compare these predictions with the classification of the sources based on the detection of an optical counterpart or secondary lines and assign them to three categories: observed transition and most probable one agree, the observed transition is one $J_{CO}$ lower/higher than the most probable one, the difference between observed and predicted is larger than two $J_{CO}$ levels.

In the case of ASPECS detections, the prediction of the highest probability transition agrees well with the actual transition for 45 per cent of the candidates. However, for most of the remaining candidates, the actual transition was the second or third most probable one. These statistics are more favourable for the PHIBSS2 detections, where 60 per cent of candidates had the highest probability transition identical to the confirmed one.  We summarised the results of the test for both surveys in Fig. \ref{fig:pie}. We stress that in our study we do not choose a single transition with the highest probability, but include all predictions in a probabilistic approach.

In general, we observe that the ASPECS and PHIBSS2 sources are brighter than the sources observed in ALMACAL-CO. Their probability function predicted from \textsc{Shark} is also steeper than the ones assigned to ALMACAL-CO sources (80-90 per cent to almost possible transitions while for ALMACAL-CO sources the maximal is $\sim$ 45 per cent). 

The calculator operates on single objects and does not include the parameters of the surveys like frequency coverage or probed volume. With the volume probed by ASPECS in CO (1$-$0) it would not be possible to observe as high a fraction of low redshift CO(1$-$0) detections as the redshift calculator suggests. It is also more probable, from the argument of the population, to detect a moderately bright object nearby (CO(1$-$0) or CO (2$-$1) $z$ < 0.5 ) than a very bright higher $J_{\rm up}$ transition at higher $z$ (like CO(3$-$2), $z \sim$ 1.5). 

\subsection{Alternative identification of sources as [CII]}
\label{sec:cplus}

With the lack of the optical counterparts associated with emission line candidates, we cannot \textit{a priori} exclude the possibility of the sample contamination by the species different from CO. Three of our detections could be associated with redshift $z \sim$ 4.5 [CII] emission. [CII] emission predictions are not included in \textsc{Shark} and are not part of the redshift calculator. 

The luminosity function for [CII] at $z \sim 4.5 $ is not well constrained observationally \citep[limits by][]{swinbank12,matsuda15}, therefore we referred to simulated [CII] luminosity function to estimate the number of expected sources in the volume covered by ALMACAL-CO pilot. We used the [CII] luminosity functions from \citet{popping16,lagache18} semi-analytical model to calculate the expected number of sources at redshift $z \sim$ 4.5 with a luminosity L = 10$^{7.5}$ L$_{\odot}$ for the probed volume. Regardless of the model used, the predicted number of [CII] sources in ALMACAL-CO pilot is less than one. Additionally, we performed similar calculations with redshift $z$ = 0 observations from Herschel by \citet{hemmati17}, obtaining the same results. As the [CII] solution is deemed unlikely, we only include the associations with CO lines in our analysis. 

\begin{figure}
	\includegraphics[width=\columnwidth]{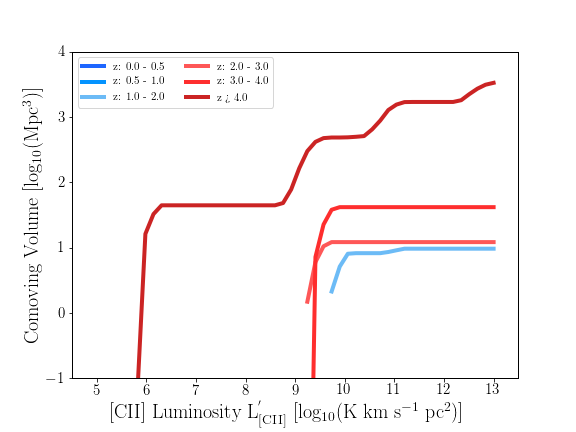}
    \caption{ALMACAL-CO pilot volume coverage for a range of line luminosities of [CII] emission line. The range of lines from blue to red represents different redshift bins as indicated in the legend. ALMACAL-CO pilot covers small volume for [CII] emission line. From the [CII] luminosity function we estimate the number of expected detections of the [CII] line in the volume probed to be less then one. We therefore assume all line detections are CO.}
    \label{fig:sledo}
\end{figure}
\label{sec:app_cand}


\section{Identification of CO lines candidates}
  
\subsection{ALMACAL search for rotation lines associated with CO candidates}

Single emission line detections in ALMACAL-CO pilot pose a challenge for line classification. However, the advantage of ALMACAL is the availability of wider frequency coverage so that potentially other CO lines could be searching for in the data not included in this pilot survey. To complement the \textsc{Shark}-based redshift probability calculator results, we looked for additional data to help us confirm and identify the detections. 
Since the ALMACAL-CO pilot consists of a subset of ALMACAL survey data, we looked into the remaining data available for the calibration fields to look for additional CO lines. For each candidate, we assume a CO transition classification and we checked the frequencies of the corresponding CO lines from CO(1$-$0) to CO(10$-$9). We centred the new cubes around the observed frequency of the given CO transitions and consider the data within $\pm$ 1000 km s$^{-1}$ around it. 

The frequency coverage of ALMACAL data differs significantly between the calibrator fields. For the ALMACAL-CO pilot fields checked, the percentage of additional transitions covered by the data ranges from 20 to 33  per cent between candidates, making the probability to detect additional transitions rather low. The resulting coverage of the additional CO transition is uneven for different candidates and cannot be used to exclude some of the classification possibilities. 

To estimate if the r.m.s of the new data was sufficient for the CO lines to be detected, we used the prediction of CO emission lines brightness form \textsc{Shark}. Each detection can be assigned a certain CO $J_{\rm up}$ transition for which we generate corresponding average simulated CO SLED. We then compare the expected brightness of the CO line with the r.m.s of the datacube covering a given transition. We check which of the simulated line brightness exceed the signal-to-noise ratio of 3 $\sigma$, to classify it as detectable. Among all covered transitions, we should be able to detect 48 per cent of the lines. The results differ in between calibrator fields, from 17 to 62 per cent of the covered transition in the field.

We searched for additional detections in these additional cubes by visually inspecting the data at the expected line frequency and we run SoFiA source finder on the cubes. With both approaches, we do not find any detections corresponding to secondary lines of ALMACAL-CO candidates. However, most of the top-probability lines predicted by \textsc{Shark} are not well covered in the available data, therefore the non-detections are not constraining.

The availability of the broad frequency coverage of the ALMACAL fields is a useful tool for providing constraints for the emission lines detection classification. The test presented here will be extended for the full ALMACAL-CO sample search and can become more constraining.


 



\bsp	
\label{lastpage}
\end{document}